\documentclass[pre,epsfig,aps,superscriptaddress,showpacs,twocolumn]{revtex4}
\usepackage{epsfig}
\usepackage{latexsym}
\usepackage{graphicx}
\usepackage{multirow}
\usepackage{threeparttable}
\usepackage{amssymb,amsfonts,amsmath}

\begin{document}
\title{Investigation on energetic optimization problems of stochastic thermodynamics with iterative dynamic programming}
\author{Linchen Gong}
\affiliation{Center for Advanced Study, Tsinghua University, Beijing, 100084, China}
\author{Ming Li\footnote{Author to whom correspondence should be addressed; Electronic Mail: liming@gucas.ac.cn}}
\affiliation{College of Physical Science, Graduate university of Chinese Academy of Sciences, Beijing, 100190, China}
\author{Zhong-can Ou-yang}
\affiliation{Center for Advanced Study, Tsinghua University, Beijing, 100084, China}
\affiliation{Institute of Theoretical Physics, Chinese Academy of Sciences, Beijing, 100190, China}

\begin{abstract}
The energetic optimization problem, {\it e.g.}, searching for the optimal switching protocol of certain system parameters to minimize the input work, has been extensively studied by stochastic thermodynamics. In current work, we study this problem numerically with iterative dynamic programming. The model systems under investigation are toy actuators consisting of spring-linked beads with loading force imposed on both ending beads. For the simplest case, {\it i.e.}, a one-spring actuator driven by tuning the stiffness of the spring, we compare the optimal control protocol of the stiffness for both the overdamped and the underdamped situations, and discuss how inertial effects alter the irreversibility of the driven process and thus modify the optimal protocol. Then, we study the systems with multiple degrees of freedom by constructing oligomer actuators, in which the harmonic interaction between the two ending beads is tuned externally. With the same rated output work, actuators of different constructions demand different minimal input work, reflecting the influence of the internal degrees of freedom on the performance of the actuators.
\end{abstract}


\maketitle

\section{Introduction}
The past two decades have seen the rapid growth of researches on nano-machines. Various types of nano-machines have been proposed and synthesized for practical purposes~\cite{huang,kay}. This progress presents an urgent appeal and also a great challenge for physicists to understand the working principle and energetics of nano-machines. For isothermal nano-machines including protein machines and their derivatives or mimics, researchers are able to formulate a conceptual framework, the Brownian ratchet, to understand in general how a {\it scalar} energy source (light, chemical reaction, thermal agitation, etc.) can facilitate a {\it vectorial} process. The ratchet mechanism has not only been used to explain many experimental observations on chemically-driven protein motors~\cite{julicher1,fisher1,oster1}, but also been successfully employed to develop some micro-manipulation techniques ({\it e.g.}, a ratchet-like micro-device was designed for DNA segregation~\cite{bader}). Within the ratchet framework, the energetics (or thermodynamics) has also been extensively investigated with multiple-state Langevin equations or Fokker-Planck equations ({\it e.g.}, see~\cite{julicher1} for chemically-driven nano-machines, and~\cite{suzuki} for externally-controllable nano-machines).

In this context, some general energetic topics of nano-machines can be put forward. Particularly, in analogy to the optimization theory based on macroscopic finite-time thermodynamics~\cite{berry1,berry2}, the energetic optimization problems can be re-formulated for ratchet-type nano-machines. For instance, one may require the {\it m}inimal {\it i}nput {\it w}ork (abbreviated as MIW) in a finite-time process to achieve least heat agitation~\cite{then} or maximum power~\cite{seifert2} or highest efficiency~\cite{astumian}. To properly formulate these questions, choosing proper model systems is very important. While the multiple-state ratchet model may not be a proper candidate (especially, not easy for analytical treatment), Sekimoto and Seifert {\it et al.} initiated the study with a much simpler ratchet model. In this model, the machine performs Brownian motion on an energy landscape. And, a few parameters that characterize the landscape can be switched externally according to a deterministic protocol~\cite{sekimoto1,seifert2}. For such driven systems, stochastic thermodynamics has been firmly established, and the work and heat can be co-identified under the construction of first-law-like and second-law-like thermodynamic relations~\cite{sekimoto1,sekimoto2,seifert1}. With the well-defined model systems and thermodynamics, a basic question is addressed, {\it i.e.}, what is the {\it o}ptimal {\it p}rotocol of the switching parameters to realize the \textit{m}inimal \textit{i}nput \textit{w}ork (abbreviated as OPMIW). For the simple cases that the Brownian motion of an overdamped or underdamped bead is controlled by a tunable harmonic potential, the authors were able to present analytical solutions for the OPMIWs (see \cite{seifert2,seifert3} for overdamped case, and \cite{seifert4} for underdamped case). While their analysis focused on analytically solvable potentials, Then and Engel also presented a Monte Carlo numerical method to discuss similar optimization problems of more complicated systems~\cite{then}.

In this work, we follow the same logic and introduce a numerical method called {\it i}terative {\it d}ynamic {\it p}rogramming (abbreviated as IDP)~\cite{luus} to investigate the OPMIW problems. For systems with polynomial potential functions up to the second order, the OPMIW problem is equivalent to the conventional optimal control problem, thus can be solved numerically with IDP. Inspired by the controllable DNA nano-actuator converting chemical energy into mechanical work~\cite{liud1,liud2}, we choose toy actuators as our model systems. The simplest construction is a two-bead actuator, in which the two beads are linked by a spring with externally tunable stiffness (a polymer can be regarded as such an actuator if its persistence length is tuned externally). Although this model has been discussed by Seifert {\it et al.}~\cite{seifert2}, the promising numerical method enables us to study the more complicated case when the actuator is working against a pulling force, and consequently to study the relation between the output work and the input work. Furthermore, since IDP can efficiently treat systems with multiple degrees of freedom~\cite{luus}, actuators with different internal structures can be constructed and the related OPMIW problems can be discussed by this method. Besides, additional constraints can be easily specified (as will be clear in subsequent sections) to study some interesting issues on energetics.

In the following, we will first present the formulation of OPMIW problems for various toy models. In section \ref{sec:result}, after testing the numerical precision of IDP for two analytically solvable examples, we apply IDP to simple systems with overdamped or underdamped dynamics, and also to systems with multiple degrees of freedom. Some general remarks will be presented in section \ref{sec:conclud}. The details of IDP method can be found in the appendix.


\section{Formulation for OPMIW problem}
In this section, we will present the mathematical formulation of OPMIW problem for various model systems. We consider the simplest model of a one-spring actuator first, and then generalize the description to cases with multiple degrees of freedom.

\subsection{Formulation for one-spring actuator}
\label{sec:simple}
\begin{figure}
\includegraphics[width=3.5in]{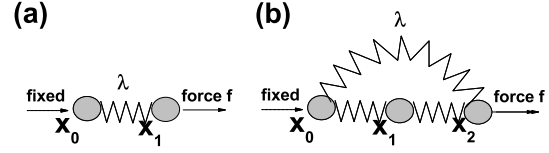}\\
\caption{Illustration of the models in study. (a) The one-spring actuator with constant pulling force $f$. (b) The oligomer actuator. The $N=2$ case is portrayed.}
\label{fig:0}
\end{figure}

The one-spring actuator is basically the same model that Seifert and colleagues have considered before, in which a bead is trapped in a stiffening harmonic potential~\cite{seifert2}. The actuator is illustrated in Fig.~\ref{fig:0}(a), where two beads are connected by a spring with externally tunable stiffness. To study the energetic problems, one bead is fixed at the origin while the other is pulled by a constant force.

We only focus on one-dimensional problems here, thus the potential energy of the one-spring actuator can be written as
\begin{equation}
\label{potential} V(x,\lambda)=\frac{1}{2}\lambda{}x^2-fx,
\end{equation}
where $\lambda$ is the stiffness of the spring, $x$ is the position of the movable bead, and $f$ is a constant pulling force (suppose $f$ is pointing to the positive direction of $x$-axis). For the overdamped case, the Langevin equation describing the motion of the movable bead is
\begin{equation}
\label{langevin} \gamma\frac{dx}{dt}=-\lambda{}x+f+\sqrt{2k_BT\gamma}\xi(t),
\end{equation}
where $\gamma$ is the frictional coefficient, $k_B$ is the Boltzmann constant, $T$ is the temperature of the thermostat, and $\xi(t)$ is Gaussian white noise satisfying
\begin{equation}
\langle \xi(t)\xi(t')\rangle=\delta(t-t').
\end{equation}
In the following, $\gamma$ and $k_BT$ are set equal to unity for simplicity. If the unit of another quantity is specified, {\it e.g.}, the force $f$, the dimensions of all the quantities involved in the model can be expressed by the dimensions of $\gamma$, $k_BT$ and $f$. The Fokker-Planck equation for the probability evolution of $x$ reads
\begin{equation}
\label{fke} \frac{\partial{}g(x,t)}{\partial{}t}=\frac{\partial{}}{\partial{}x}\left[g(x,t)\frac{\partial{}V(x,\lambda)}{\partial{}x}+
\frac{\partial{}g(x,t)}{\partial{}x}\right],
\end{equation}
where $g(x,t)$ is the probability distribution function of $x$ at time $t$.

According to stochastic thermodynamics~\cite{seifert2}, under the framework of Langevin equation [Eq.~(\ref{langevin})], the work increment $dw$ along the stochastic trajectory of $x$ is
\begin{equation}
d{}w=\frac{\partial{}V}{\partial{}\lambda}\dot{\lambda}dt=\frac{1}{2}\dot{\lambda}x^2dt.
\end{equation}
Averaging with the Fokker-Planck equation [Eq.~(\ref{fke})], when $\lambda$ is switched from an initial value $\lambda_i$ to a final value $\lambda_f$ within finite time, the expected input work $W_{in}$ to the actuator can be expressed as
\begin{equation}
\label{optimal} W_{in}=\int_{t_i}^{t_f}\frac{1}{2}\dot{\lambda}\left\langle{}x^2\right\rangle{}dt,
\end{equation}
where $t_i$ and $t_f$ are the beginning and ending time point of the switching process respectively. The dot over $\lambda$ denotes the time derivative of $\lambda$. The angular brackets denote an instant ensemble average using the probability distribution $g(x,t)$. During the switching process, the average position $\langle x \rangle$ is changing, consequently the finite-time output work of the actuator is defined as
\begin{equation}
W_{out}=f\langle x \rangle|_{t=t_i} - f\langle x \rangle|_{t=t_f}.
\end{equation}
There exist several differential constraints on the moments of $g(x,t)$ when optimizing $W_{in}$. Multiplying both sides of Eq.~(\ref{fke}) by $x$ or $x^2$ and taking average, we obtain the following relations
\begin{equation}
\label{cons1} \dot{\left\langle{}x\right\rangle}=f-\lambda\left<x\right>,
\end{equation}
\begin{equation}
\label{cons2} \dot{\left\langle{}x^2\right\rangle}=2-2\lambda\left<x^2\right>+2f\left<x\right>.
\end{equation}
In IDP calculation, we usually consider $W_{in}$ as a function of time by defining
\begin{equation}
W_{in}(t)=\int_{t_i}^{t}\frac{1}{2}\dot{\lambda}\left\langle{}x^2\right\rangle{}dt,
\end{equation}
which satisfies the following differential equation
\begin{equation}
\label{wderiv}
\dot{W}_{in}=\frac{1}{2}u\langle x^2 \rangle.
\end{equation}
Here, the time derivative of $\lambda$-protocol has been explicitly expressed as a temporal function $u$, {\it i.e.},
\begin{equation}
\label{lamderiv}
\dot{\lambda}=u.
\end{equation}
Eq.~(\ref{wderiv}) and Eq.~(\ref{lamderiv}), together with Eq.~(\ref{cons1}) and Eq.~(\ref{cons2}), compose the formulation of the OPMIW problem for current system. The task now turns into searching for the optimal trajectory of $u$ that minimizes $W_{in}(t_f)$. Once $t_i$, $t_f$, $\lambda_i$, $\lambda_f$ and the initial values of $\langle{}x\rangle$ and $\langle{}x^2\rangle$ are all specified [$W_{in}(0)$ is zero by definition], the problem can be solved with IDP method.

For underdamped one-spring actuator, while $W_{in}$ is the same with Eq.~(\ref{optimal}), the set of differential constraints become
\begin{equation}
\dot{\langle x^2\rangle }=2\langle xp\rangle/m,
\end{equation}
\begin{equation}
\dot{\langle xp\rangle }=\langle p^2\rangle/m-\langle xp\rangle/m-\lambda \langle x^2\rangle + f\langle x\rangle,
\end{equation}
\begin{equation}
\dot{\langle p^2\rangle }=-2\langle p^2\rangle/m-2\lambda \langle xp\rangle + 2f\langle p\rangle+2,
\end{equation}
\begin{equation}
\dot{\langle x\rangle }=\langle p\rangle/m,
\end{equation}
\begin{equation}
\dot{\langle p\rangle }=-\langle p\rangle/m -\lambda \langle x\rangle + f.
\end{equation}
Where, $p$ and $m$ are the momentum and mass of the movable bead, respectively.

\subsection{Generalization to models with multiple degrees of freedom}
In this subsection, we consider a nano-system with coordinates vector $\vec{x}$ and time-dependent potential function $V(\vec{x},\vec{\lambda}(t))$. The parameter vector, $\vec{\lambda}$, of the potential is externally tuned according to certain protocol. To model the actuation process, constant forces are loaded on the system, thus the potential function can be written as
\begin{equation}
\label{potential_force} V(\vec{x},\vec{\lambda})=V_0(\vec{x},\vec{\lambda})-\vec{f}\cdot\vec{x},
\end{equation}
where, $V_0(\vec{x},\vec{\lambda})$ represents the internal energy of the system ({\it e.g.}, the elastic energy of the springs in the toy actuator). $\vec{f}$ is not a normal three-dimensional force vector, instead, it is a high-dimensional vector with each element representing the force on the corresponding degree of freedom.

The dynamics of $\vec{x}$ can be described by overdamped or underdamped Langevin equations. For example, in overdamped case, the Langevin equations read
\begin{equation}
\label{langevin2} \frac{dx_j}{dt}=-\frac{\partial{}V(\vec{x},\vec{\lambda})}{\partial{}x_j}+\sqrt{2}\xi_j(t),\quad{}j=1,2,\ldots,d,
\end{equation}
where $d$ is the dimension of $\vec{x}$. For simplicity, we suppose that the random forces on different elements of $\vec{x}$ are uncorrelated, and the frictional coefficients are all the same. Thus, the noise terms $\{\xi_j\}$~($j=1,2,\ldots,d$) satisfy
\begin{equation}
\langle \xi_i(t)\xi_j(t')\rangle=\delta(t-t')\delta_{ij}.
\end{equation}
It is worth noting that models without such restrictions can also be solved with IDP.

In the same spirit as section \ref{sec:simple}, when $\vec{\lambda}$ is switched from its initial value $\vec{\lambda}_i$ to the final value $\vec{\lambda}_f$ in finite time, the input work to the system, $W_{in}$, can be formally expressed as
\begin{equation}
\label{work_input} W_{in}=\int_{t_i}^{t_f}\dot{\vec{\lambda}}\cdot\left\langle\nabla_{\lambda}V(\vec{x},\vec{\lambda})\right\rangle{}dt
\end{equation}
where $\nabla_{\lambda}$ is the gradient with respect to $\vec{\lambda}$, and the brackets again denotes the average over the instant distribution function $g(\vec{x},t)$ of the system. Besides, during the switching, the change of $\langle \vec{x} \rangle$ leads to an output work
\begin{equation}
W_{out}=\vec{f}\cdot\langle \vec{x} \rangle|_{t=t_i}-\vec{f}\cdot\langle \vec{x} \rangle|_{t=t_f}.
\end{equation}
If $V(\vec{x},\vec{\lambda})$ is a polynomial function up to the second order in $\vec{x}$, the number of differential constraints on the moments of $g(\vec{x},t)$ would be finite, the OPMIW problem can be rigorously expressed as an IDP-sovable optimization problem.

To better illustrate the energetics of OPMIW, we only focus on the one-way switching processes of $\vec{\lambda}$ throughout this article. One can integrate these and other processes into a complete working cycle to construct a novel nano-machine, as illustrated in Ref.~\cite{seifert3}, the related optimization problem may need to be re-formulated. Besides, in present study, the systems are initially equilibrated under the potential function $V(\vec{x},\vec{\lambda_{i}})$. Consequently, the initial conditions of the differential constraints, {\it i.e.}, the initial values of the moments of $g(\vec{x},t)$, are selected as the corresponding equilibrium values. However, both the initial and final values of the moments can be chosen arbitrarily in computation as long as the optimal protocol exists. Setting the final values of the moments of $g(\vec{x},t)$ means that additional constraints are included in OPMIW problem.

\subsection{Formulation for oligomer actuators}
\label{sec:complex}
Under the general framework for systems with multiple degrees of freedom, we consider the one-dimensional oligomer actuators consisting of $N+1 (N=1,2,...)$ beads in this subsection. Instead of the linear oligomers without branches and cross-links, we consider a sub-family of structured actuators with the potential function
\begin{equation}
\label{multpot}
V(x_1,x_2,...,x_N)=\frac{1}{2}x_1^2+\frac{1}{2}\Sigma_{i=2}^N(x_i-x_{i-1})^2+\frac{1}{2}\lambda{}x_N^2-fx_N.
\end{equation}
Here, the beads are numbered from $0$ to $N$, $x_i$ denotes the position of the $i$th bead. While the adjacent beads are connected by a spring with elastic coefficient $1.0$, $x_N$ is connected to $x_0$ by an additional spring with tunable elastic coefficient $\lambda$. To study the energetic problems, $x_0$ is assumed to be fixed at the origin, a pulling force $f$ is imposed on $x_N$. The actuators are driven by switching $\lambda$ from the initial value $\lambda_i$ to the final value $\lambda_f$ in finite-time. The current model system is inspired by the recently designed DNA oligomer actuators~\cite{liud1, liud2}. In these DNA actuators, certain intra-molecular bonds can be modified by switching the light or chemical conditions, which results in considerable change of end-to-end distance ({\it i.e.}, a potential actuation process). Here, in the same spirit, a model actuator achieves an actuation when the 'intra-molecular bond' between bead $0$ and bead $N$ is tuned externally. Therefore, the $N=1$ model is exactly the above-mentioned one-spring actuator in which the only bond is tuned, and thus only the last two energy terms in Eq.~(\ref{multpot}) need to be included in computation. The $N>1$ models describe structured actuators with extra degrees of freedom, {\it e.g.}, the $N=2$ actuator illustrated in Fig.~\ref{fig:0}(b).

In current article, overdamped Langevin equations are chosen to model the dynamics of the actuators. For $N=2$ case, the optimal index, {\it i.e.}, the input work, $W_{in}$, is
\begin{equation}
{}W_{in}=\int_{t_i}^{t_f}\frac{1}{2}\dot{\lambda}\langle x_2^2\rangle dt.
\end{equation}
The differential constraints are
\begin{equation}
\dot{\langle x_1^2\rangle}=-4\langle x_1^2\rangle+2\langle x_1x_2\rangle+2,
\end{equation}
\begin{equation}
\dot{\langle x_1x_2\rangle}=-(3+\lambda)\langle x_1x_2\rangle+\langle x_1^2\rangle+\langle x_2^2\rangle+f\langle x_1\rangle,
\end{equation}
\begin{equation}
\dot{\langle x_2^2\rangle}=-2(1+\lambda)\langle x_2^2\rangle+2\langle x_1x_2\rangle+2f\langle x_2\rangle+2,
\end{equation}
\begin{equation}
\dot{\langle x_1\rangle}=-2\langle x_1\rangle+\langle x_2\rangle,
\end{equation}
\begin{equation}
\dot{\langle x_2\rangle}=-(1+\lambda)\langle x_2\rangle+\langle x_1\rangle+f.
\end{equation}

For $N=3$ case, $W_{in}$ is
\begin{equation}
{}W_{in}=\int_{t_i}^{t_f}\frac{1}{2}\dot{\lambda}\langle x_3^2\rangle dt.
\end{equation}
The differential constraints are
\begin{equation}
\dot{\langle x_1^2\rangle }=-4\langle x_1^2\rangle +2\langle x_1x_2\rangle +2,
\end{equation}
\begin{equation}
\dot{\langle x_2^2\rangle }=-4\langle x_2^2\rangle +2\langle x_1x_2\rangle +2\langle x_2x_3\rangle +2,
\end{equation}
\begin{equation}
\dot{\langle x_3^2\rangle }=-2(1+\lambda)\langle x_3^2\rangle +2\langle x_2x_3\rangle +2f\langle x_3\rangle +2,
\end{equation}
\begin{equation}
\dot{\langle x_1x_2\rangle }=-4\langle x_1x_2\rangle +\langle x_2^2\rangle +\langle x_1^2\rangle +\langle x_1x_3\rangle,
\end{equation}
\begin{equation}
\dot{\langle x_2x_3\rangle }=-(3+\lambda)\langle x_2x_3\rangle +\langle x_1x_3\rangle +\langle x_3^2\rangle +\langle x_2^2\rangle +f\langle
x_2\rangle,
\end{equation}
\begin{equation}
\dot{\langle x_1x_3\rangle }=-(3+\lambda)\langle x_1x_3\rangle +\langle x_2x_3\rangle +\langle x_1x_2\rangle +f\langle x_1\rangle,
\end{equation}
\begin{equation}
\dot{\langle x_1\rangle }=-2\langle x_1\rangle +\langle x_2\rangle,
\end{equation}
\begin{equation}
\dot{\langle x_2\rangle }=-2\langle x_2\rangle +\langle x_1\rangle +\langle x_3\rangle,
\end{equation}
\begin{equation}
\dot{\langle x_3\rangle }=-(1+\lambda)\langle x_3\rangle +\langle x_2\rangle +f.
\end{equation}

\section{Numerical results and Discussions}
\label{sec:result}
\subsection{Numerical accuracy of IDP}
\label{sec:numtest}
To test the numerical accuracy of IDP, we first investigate the two examples that have been analytically solved by Schmiedl and Seifert~\cite{seifert2}. Both of the two systems are one-dimensional systems satisfying overdamped Langevin dynamics. The two model potentials are $V_1(x, \lambda)=\frac{1}{2}(x-\lambda)^2$ (with tunable average position) and $V_2(x, \lambda)=\frac{1}{2}\lambda{}x^2$ (with tunable stiffness) respectively.

\begin{figure}
\includegraphics[width=3.5in]{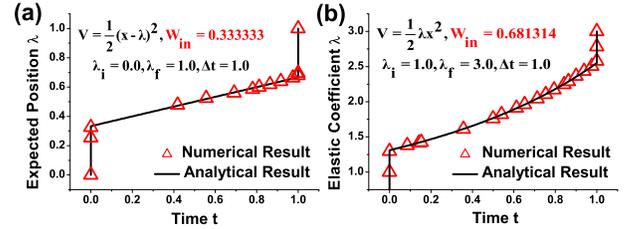}\\
\caption{(color online). Comparison between numerical and analytical solutions [Eq.(9) and Eq.(19) in \cite{seifert2}] for overdamped systems with harmonic potentials. (a) Results for the system with tunable averaged position $\lambda$. (b) Results for the system with tunable elastic constant (spring stiffness) $\lambda$. The potential functions and related parameters are shown in the figure.}
\label{fig:1}
\end{figure}

The comparison between the numerical results and the analytical solutions is shown in Fig.~\ref{fig:1}, where the potential functions, the initial and final values of $\lambda$, the switching time interval $\Delta{t}$ and the MIW, $W_{in}$, are all listed. We only list one MIW value in either figure, since the numerical MIW is the same with the analytical MIW [Eq.~(9) and Eq.~(19) in~\cite{seifert2}] up to the shown numerical accuracy. In either figure, the solid line is the analytical protocol of $\lambda$, and the hollow triangles denote the numerical results. Obviously, for the two examples studied here, the numerical results accurately follow the analytical curves. The important characteristics of the OPMIWs for the model systems, {\it i.e.}, the jumps both at the beginning and at the end of the optimal protocols, are successfully reproduced.

The triangles in Fig.~\ref{fig:1} distribute non-uniformly along the time axis due to the employed method of variable step-length. In this scheme, the space between neighboring time points are also optimized. Hence, more time points will be gathered at the sharply changing parts of the $\lambda$-protocol, with less at the smoothly changing parts. As a result, higher accuracy can be obtained with limited discretization of $[t_i, t_f]$. Implementing the variable step-length method does not change the procedure of IDP (see the appendix).

\subsection{OPMIW for overdamped cases}
In this subsection, we consider the overdamped one-spring actuator introduced in section \ref{sec:simple}. Without the pulling force, this model turns into the analytical solvable model with tunable stiffness (see section \ref{sec:numtest}). The force complicates the problem, making it very hard to get an analytical solution. However, the model system is quite suitable for IDP. For all the processes studied in this subsection, $\lambda$ is switched from $1.0$ to $3.0$, and the force is set as $1.0$ if not further claimed.

\begin{figure}
\includegraphics[width=3.5in]{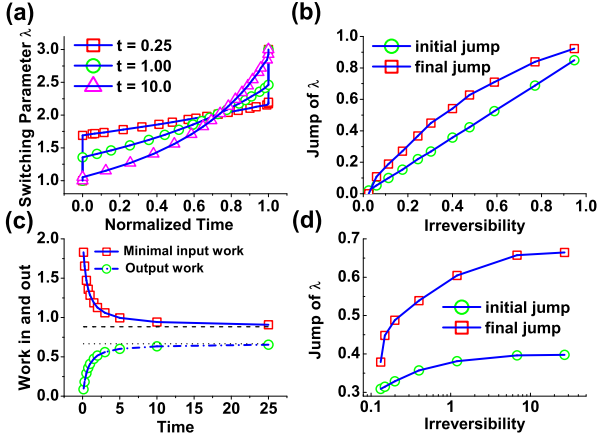}\\
\caption{(color online). Numerical results for the overdamped one-spring actuator. (a) The optimal protocols with switching time $0.25$ (squares), $1.00$ (circles) and $10.0$ (upward triangles). The switching time of the protocols have been re-scaled to $[0.0,1.0]$ for better visualization. The stiffness of the spring, $\lambda$, is switched from $1.0$ to $3.0$. (b) Relation between the initial (circles) or final (squares) jump in OPMIW and the irreversibility, $W_{in}-\Delta{F}$. The data points correspond respectively to processes with switching times of $25.0$, $10.0$, $5.0$, $3.0$, $2.0$, $1.5$, $1.0$, $0.75$, $0.50$, $0.25$ and $0.10$ (in the order of increasing irreversibility). (c) The minimal input works (squares) and the corresponding output works (circles) for processes with different switching time. The horizontal dashed line indicates the free energy difference between the two equilibrium states determined by $\lambda_i$ and $\lambda_f$. The horizontal dotted line indicates the maximal output work of the system, which can only be realized with infinite switching time. (d) Relation between the initial or final jump in OPMIW and the irreversibility, $W_{in}-\Delta{F}$. The data points correspond respectively to processes with different pulling forces of $0$, $0.25$, $0.5$, $1.0$, $2.0$, $5.0$ and $10.0$ (in the order of increasing irreversibility, the switching time is set as $1.0$).}
\label{fig:2}
\end{figure}

We first solve the OPMIW problems for processes with different switching time intervals and the same $\lambda_i$, $\lambda_f$ and $f$. Some of the OPMIWs are shown in Fig.~\ref{fig:2}(a). As can be seen, there are always jumps at the beginning and at the end of the optimal protocols, which is similar to the cases in the already studied systems~\cite{then,seifert3}. Intuitively, with infinite or large enough time interval, all the jumps should approach zero no matter how large the force is (or even whatever the potential function is), because the switching process is actually quasi-static or reversible. 'reversible' here means that the input work can be completely converted into the free energy of the system, without any heat dissipation. This is consistent with the second law of stochastic thermodynamics $W_{in}\geq\Delta{F}$, where $\Delta{F}$ is the free energy difference between the two equilibrium states that are determined by $\lambda_i$ and $\lambda_f$ respectively. This inequality holds for any protocols of control variables in a driven process including the OPMIW. To verify this, the MIWs for processes with different switching time are shown in Fig.~\ref{fig:2}(c). It can be seen that the MIW decreases monotonically with the increase of switching time, and asymptotically reaches $\Delta{F}$ when the switching time goes to infinity [the value of $\Delta{F}$ is indicated as the dashed line in Fig.~\ref{fig:2}(c)]. Consequently, the non-negative quantity $W_{in}-\Delta{F}$ can be taken as an indicator of the irreversibility of a driven process. In fact, it is exactly the total dissipation of the actuator and the environment~\cite{sekimoto1}. In contrast to the input work, as shown in Fig.~\ref{fig:2}(c), the output mechanical work increases monotonically with the switching time and also reaches an asymptotic value that is only determined by the initial and final equilibrium states. When the system finally reaches equilibrium, the change of average position $\langle x \rangle$ reaches its maximum with the largest output work. Thus, longer switching time results in a final state closer to equilibrium and a larger output work. The above results suggest that IDP can provide physically sound and numerically correct results for the OPMIW problems discussed here.

While the infinitely slow process without any jumps leads to the lower bound (zero) of dissipation, the upper bound of dissipation, {\it i.e.}, the highest irreversibility, occurs in the case of zero switching time ({\it i.e.}, $\lambda$ is switched from $\lambda_i$ to $\lambda_f$ instantaneously). Between these two extremes are the processes with finite switching time. Thus, it is reasonable to expect that the shorter switching time corresponds to the larger jumps in OPMIW [see Fig.~\ref{fig:2}(a)] due to the increased irreversibility. For switching processes with different time intervals, the relations between the magnitude of jumps (both of the initial and of the final jumps) and the irreversibility of the OPMIW are plotted in Fig.~\ref{fig:2}(b). The trends of these curves indeed conform to the above guess. We also notice that when the switching time approaches zero, {\it i.e.}, almost with the largest irreversibility, the OPMIW becomes stepwise and has very little advantage over the other control protocols: different $\lambda$-protocols result in almost the same input work since the switching is too fast to change the position of the bead.

Similarly, we also study the processes with the same switching time interval and various pulling forces. For current model system, larger force results in larger irreversibility, which, as expected, leads to larger jumps in the optimized $\lambda$-protocol [see Fig.~\ref{fig:2}(d)]. From Fig.~\ref{fig:2}(b) and (d), together with other results of changing the switching range of $\lambda$ (data not shown) or the bead mass (the inertial effects will be discussed in the next subsection), we expect that the magnitude of either the initial or the final jump in OPMIW may always monotonically correlate to the irreversibility, no matter what the potential function is. Of course, this observation needs further validation in other systems.

\subsection{OPMIW for underdamped cases}
In this subsection, we study the underdamped one-spring actuator introduced in section \ref{sec:simple}. Similar to the overdamped cases, the OPMIWs for the underdamped actuators also have initial and final jumps. Besides, for the system with tunable average position in section \ref{sec:numtest}, $\delta$-function-like pulses have been analytically found in the optimized $\lambda$-protocol. Although the current model system can not be solved analytically, the $\delta$-function-like pulses in OPMIW have also been inferred when the pulling force is zero~\cite{seifert4}. For all the processes studied in this subsection, a constant force $f=1.0$ is always loaded on the movable bead, $\lambda$ is switched from $1.0$ to $3.0$ with switching time $1.0$.

\begin{figure}
\includegraphics[width=3.5in]{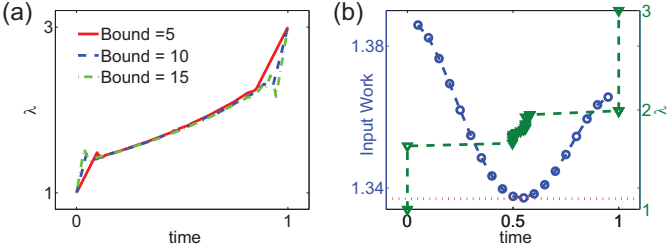}\\
\caption{(color online). The OPMIWs of underdamped one-spring actuator. (a) The non-monotonic OPMIWs for the $m=0.02$ actuator with different bounds of $u\equiv\left|\frac{d\lambda}{dt}\right|$. The results obtained with the bound of $5$ (solid line), $10$ (dashed line) and $15$ (dash-dotted line) are shown. (b) The input works for protocols with three jumps and two plateaus ($m=0.2$, see the main text for more details). The horizontal axis indicates the position of the intermediate jump. The left vertical axis shows the input work. The OPMIW (downward triangles) for this model is also shown for comparison. The right vertical axis shows the $\lambda$ value. The dotted horizontal line indicates the MIW value for this model.}
\label{fig:3}
\end{figure}

We first calculate the OPMIW for the one-spring actuator with mass $m=0.02$. Three optimal protocols with different allowable range of $u\equiv\frac{d\lambda}{dt}$ are plotted in Fig.~\ref{fig:3}(a). In contrast to the overdamped cases, the optimized $\lambda$-protocols are no longer monotonic. There are always an upward peak at the beginning and a downward peak at the end of the OPMIW. The upward (downward) slope of either peak always matches the upper (lower) bound of $u$, indicating that the $\delta$-function-like traces do exist in the OPMIW of current model system. The initial upward $\delta$-peak enables an instant acceleration of the bead to ensure a nearly constant velocity during the process, and the final downward $\delta$-peak enables an instant deceleration of the bead to ensure minimal dissipation~\cite{seifert4}. Nevertheless, to exactly reproduce a $\delta$-function trace of $\lambda$, the allowable range of $u$ should be very large in computation, which is numerically hard (see the discussion in appendix) and is not the focus of this article.

\begin{figure}
\includegraphics[width=3.5in]{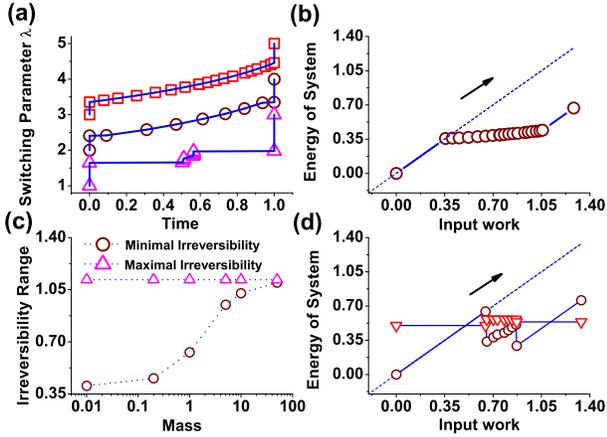}\\
\caption{(color online). Numerical results for the underdamped one-spring actuators. (a) The OPMIWs with different bead mass. The OPMIWs for $m=0.01$ (hollow circles) and $m=0.2$ (hollow triangles) are shown. For comparison, the OPMIW for the overdamped case is also shown (hollow squares). The curves have been shifted for better visualization. The relations between the instant energy terms of the system and the cumulative input work are shown in (b) (overdamped case) and (d) ($m=0.2$), where the instant potential energy is denoted by hollow circles, the cumulative input work is denoted by dotted line, the kinetic energy is denoted by hollow downward triangles. The horizontal axis is the cumulative input work. The arrows in (b) and (d) indicate the direction of time evolution. (c) shows the variation range of irreversibility. The maximum and minimum irreversibility are plotted for different bead masses of $0.01$, $0.5$, $1.0$, $5.0$, $10.0$ and $50.0$.}
\label{fig:4}
\end{figure}

To demonstrate the feasibility of IDP-based OPMIW study for underdamped systems, we turn to a subset of protocols in which $\lambda$ grows monotonically with time. This subset of protocols might be easier to implement in practice. We calculate the OPMIWs for actuators with different bead mass $m$, with two OPMIWs ($m=0.2$ and $m=0.01$) illustrated in Fig.~\ref{fig:4}(a). The initial and final jump of $\lambda$ value are found again in the OPMIWs. As expected, when $m$ decreases, thus $\gamma/m$ increases, the underdamped protocol will converge to the overdamped one. For finite $m$, however, plateaus appear immediately after the initial jump and before the final jump in an optimized $\lambda$-protocol. This phenomenon can be roughly explained as following. On one hand, after the first plateau stage, the bead has been accelerated to finite velocity. This velocity remains almost constant when $\lambda$ starts to increase again until reaching the other plateau of $\lambda$-protocol. Therefore, from the perspective of minimal dissipation, the dissipation due to non-uniform velocity is minimized by including the first plateau stage. On the other hand, from the perspective of minimal input work, in the plateau stages with almost constant $\lambda$, no work is input into the system. Besides, in the final plateau stage, the established velocity of the bead is damped. Thus, the extra dissipation after switching, which would be payed by the input work, is reduced. In a word, the protocol between the initial and final jumps can be well interpreted either from the perspective of minimal dissipation or minimal input work. In the following, we will perform more detailed analysis on the numerical OPMIWs to provide in-depth perspective of inertial effect, as well as to further verify the correctness of the computational results.

As shown in Fig.~\ref{fig:4}(a), the OPMIW becomes stepwise when $m$ increases. Taking the $m=0.2$ actuator as example, except for the initial and final jump, a narrow region with sharply increasing $\lambda$ value exists approximately in the middle of the OPMIW. Actually, this short piece of increasing $\lambda$-protocol can be well approximated by a jump of $\lambda$ value without prominent increase of input work. In contrast, the position of this narrow region is more prone to changing the input work. To illustrate this point, we compare different stepwise protocols with three jumps and two plateau stages of $\lambda$ value. The initial and final jumps of these protocols are set the same with the OPMIW of $m=0.2$ actuator (thus the magnitude of the intermediate jump is also determined), only the position of the intermediate jump is variable. For one-spring actuator with bead mass $0.2$, the input works of these protocols are plotted in Fig.~\ref{fig:3}(b). As can be seen, the stepwise protocol with the smallest input work has its intermediate jump located within the intermediate increasing region of OPMIW, and its input work is only slightly larger than the MIW. This result strongly supports that the OPMIW obtained here is an optimum of the 'protocol sub-space' ({\it i.e.}, the space of monotonical $\lambda$-protocols). When $m$ further increases, the input works of various $\lambda$-protocols will gradually converge to the same value. This can be seen from Fig.~\ref{fig:4}(c), where the possible variation range of irreversibility is shown for actuators with different mass. The variation range is defined as the maximal irreversibility minus the minimal irreversibility (or equivalently, the maximal input work minus the minimal input work). The former is obtained by instantaneous switching and is obviously identical for actuators with different mass, and the latter is obtained with OPMIW. Fig.~\ref{fig:4}(c) shows that the larger bead mass corresponds to the larger minimal irreversibility and thus the smaller variation range, which means that all the $\lambda$-protocols (including OPMIW) lead to nearly the same irreversibility and input work when $m$ approaches infinity.

The inertial effect can also be perceived from the relation between input energy and system energy. For overdamped case, the potential energy of the system [{\it i.e.}, the instant ensemble average of Eq.~(\ref{potential})] increases monotonically with the cumulative input work as shown in Fig.~\ref{fig:4}(b). By contrast, as shown in Fig.~\ref{fig:4}(d), this relation is no longer monotonic for the underdamped cases (only the $m=0.2$ case is shown here). In the first plateau stage of OPMIW, while no work is input, part of the potential energy transforms into kinetic energy and heat, which is indicated by the first downward jump on the potential energy curve and the first upward jump on the kinetic energy curve in Fig.~\ref{fig:4}(d). During the ascending stage in the middle of OPMIW, the kinetic energy remains almost constant, and the potential energy begins to increase with the cumulative input work. In the second plateau stage, both potential and kinetic energy are dissipated as heat, which is indicated by the second (downward) jump on the kinetic and potential energy curves. When $m$ becomes larger, the second jump on either energy curve disappears first and then does the first jump (data not shown). When $m$ approaches infinity, the bead is too heavy to be accelerated. Consequently, the whole system seems to be switched instantaneously from an equilibrium state to a non-equilibrium state. The input work is completely stored as potential energy and the kinetic energy remains unchanged.

\subsection{Oligomer actuator: OPMIW for more degrees of freedom }
In this subsection, we will investigate how the internal degrees of freedom can change the OPMIW and performance of an actuator. Concretely speaking, we try to study an optimal design problem, {\it i.e.}, can we construct a more efficient actuator that output the specified amount of mechanical work with least input work? To explore this question, we study the model systems introduced in section \ref{sec:complex}, with $N$ equals to $1$, $2$ and $3$ respectively. In a fast enough actuation process, an actuator might be driven out of equilibrium, thus its internal structure may begin to substantially affect the performance of the energetics, which leaves room for optimal design.

Now, we can explicitly address the optimization problem. With the same variation range of $\lambda\in[\lambda_i, \lambda_f]$, the same time interval $t\in[t_i, t_f]$ and identical pulling force, as well as an extra requirement of equal amount of output work ({\it i.e.}, rated output work), the MIW values for different structured actuators can be compared. The one with the lowest MIW should outweigh the others. For illustration, we only consider the overdamped $N=1,2,3$ models. In all the examples, $\lambda$ is switched from $1.0$ to $3.0$, the switching time interval is $[0.0, 1.0]$, the pulling force is set as $1.0$, and the rated output work is chosen as $0.3$.

\begin{figure}
\includegraphics[width=3.5in]{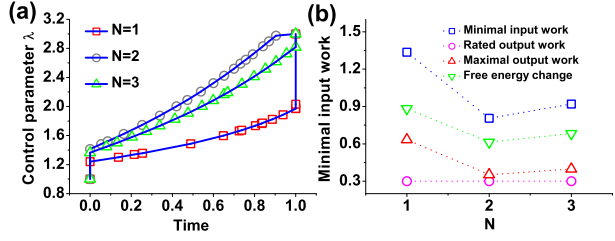}\\
\caption{(color online). Numerical results for oligomer actuators. (a) The OPMIWs for the $N=1$ (squares), $N=2$ (circles) and $N=3$ (upward triangles) models. The switching time is $1.0$, and $\lambda$ is switched from $1.0$ to $3.0$. (b) The minimal input works (squares), the maximal output works in finite-time (upward triangles), the rated output work (circles), and the free energy differences (downward triangles) for the $N=1$, $N=2$ and $N=3$ actuators.}
\label{fig:5}
\end{figure}

The OPMIWs for the $N=1,2,3$ models are plotted in Fig.~\ref{fig:5}(a). While there are both initial and final jumps in the OPMIWs of the $N=1$ and $N=3$ models, the final jump in the OPMIW of the $N=2$ actuator disappears, and a plateau region appears instead. This is a consequence of constraining the output work. As shown in Fig.~\ref{fig:5}(b), among the three models considered here, the $N=2$ actuator possesses a maximal finite-time output work which is the closest to the rated output work. Since this maximal output work is realized by instantaneously switching $\lambda$ from $\lambda_i$ to $\lambda_f$ at the beginning of the time interval (which is an infinitely fast process), the switching process of the $N=2$ actuator may also be relatively fast to realize the rated output work. Consequently, in contrast to protocols with final jump, $\lambda_f$ is reached before $t_f$ in the OPMIW of $N=2$ actuator. Actually, for any model system studied in this subsection, the OPMIW will become one-step-like when the rated output work approaches the maximal finite-time output work.

In spite of the fast switching OPMIW, the MIW of the $N=2$ model turns out to be the smallest among the three models [see Fig.~\ref{fig:5}(b)]. In other words, the $N=2$ actuator performs the best when generating a finite-time output work of $0.3$, which demonstrates the possibility for an optimal design of oligomer actuators. Interestingly, the $N=2$ model also has the smallest irreversibility among the three models [see the gap between MIW and free energy change in Fig.~\ref{fig:5}(b)]. This observation inspires us to propose an premature conjecture that irreversibility and performance might be related for actuators with similar construction and different number of degrees of freedom. If such a relation does exist, the above discussion about optimal design could be generalized no matter the intra-molecular interaction is harmonic or not (for a real macromolecule, the intramolecular interaction is usually not harmonic), since it is the irreversibility that determines the energetic performance of an actuation device.


\section{Conclusion and perspective}
\label{sec:conclud}
In current article, we study the OPMIW problems by introducing a numerical method, iterative dynamic programming (IDP). We first apply IDP to the systems with one degree of freedom and simple potential functions. Both the overdamped and the underdamped systems are investigated. For these two situations, jumps of the control variable are always found both at the beginning and at the end of OPMIW. For the underdamped case, the $\delta$-function-like traces in OPMIW are also found. With vanishing mass, the underdamped models indeed converge to the overdamped one, and with increasing mass, the OPMIW of the underdamped models become stepwise. However, with quite large mass, the OPMIW only marginally outweigh the other protocols in terms of the input work, which is similar to other cases (such as vanishing switching time) with large irreversibility.

Furthermore, we use IDP to study the optimal control and optimal design problems for the systems with multiple degrees of freedom. We consider a family of toy oligomer actuators in which the interaction between the two ending monomers can be tuned externally. The energetic performance of such actuators with different monomer number are compared, in the sense of outputting the same amount of mechanical work with as-least-as-possible input work. Our results indicate that properly choosing the number of degrees of freedom can indeed bring better energetic performance.

The above logic can be generalized to systems with more complex internal structures, {\it e.g.}, a spring network with different harmonic interactions between the bead pairs. This construction is in the same spirit with the Gaussian Network Model (GNM) widely used to describe the thermal fluctuations of folded proteins~\cite{erman} or the Elastic Network Model(ENM) used to describe the large-scale motion of motor proteins~\cite{zheng}. While those works focus on the structure-function and dynamics-function relations of protein machines, our studies might suggest a new perspective on energetics-structure relations. For instance, if the binding or release of substrates can be understood as tuning the local interaction between the residues of the motor proteins, energetic optimization problems similar to those studied here could also be raised. To carry out such investigation, however, all the thermodynamic relations and optimization problems should be re-formulated for chemically-driven protein machines. The stochastic thermodynamics of chemical reaction networks~\cite{seifert5} and the structural details of the macro-molecules should be combined, leading to a possibly formidable framework (one potential candidate is the generic ratchet model of chemically-driven machines~\cite{julicher1}) either for analytical or for numerical treatments. Therefore, at this preliminary stage, it may be beneficial to first construct some simple toy models to mimic the nano-machines. The toy actuators studied in this article can serve as the basic elements.

\appendix
\section{Iterative Dynamic Programming}
In the main text, we have presented the numerical results of IDP for various OPMIW problems. Here we provide a brief introduction to IDP. More details can be found in Ref.~\cite{luus}.

Generally speaking, IDP can solve the optimal control problems of following style.
\begin{equation}
J=\int_{t_i}^{t_f}L(\vec{s},\vec{u},t)dt,
\end{equation}
\begin{equation}
\label{prob} \dot{\vec{s}}=r(\vec{s},\vec{u},t).
\end{equation}
where $L$ is a function of the state variables $\vec{s}$, the control variables $\vec{u}$ and time $t$. Since $\vec{s}$ are determined by $\vec{u}$ through the differential constraints Eq.~(\ref{prob}), $J$ is simply an objective functional (also named as optimization index) of the protocol $\vec{u}(t)$. The related optimization problem is to find the optimal trajectory $\vec{u}(t)$ that minimizes or maximizes $J$. Once the initial condition of the state variables and the initial time $t_i$ are specified, the problem is well formulated. In principle, it is also possible to include more constraints on the final values of the state variables, the final time $t_f$ and the control variables. Conventionally, this kind of problems is solved with Pontryagin principle~\cite{pontryagin}, though it is usually hard to obtain analytical solutions. The dynamic programming method (DP)~\cite{bellman} is a conceptually sound numerical method for the optimal control problems. However, it also suffers from the difficulty of numerical interpolation and the restriction on dimensionality. In view of this, Luus developed IDP method to complement DP. With dramatic enhancement of speed, IDP is still promising for finding the global minimum, and achieves good numerical precision as shown in current article and also in Ref.~\cite{luus}. Here, we only describe the main procedures of IDP, readers can learn the underlying principles from Ref.~\cite{luus}.

By introducing an additional state variable $s^*$ satisfying
\begin{equation}
\dot{s^*}=L(\vec{s},\vec{u},t),
\end{equation}
\begin{equation}
s^*(t_i)=0,
\end{equation}
\begin{equation}
s^*(t_f)=J,
\end{equation}
the problem can be transformed into a set of differential equations shown in Eq.~(\ref{newprob1}) to Eq.~(\ref{newprob3}).
\begin{equation}
\label{newprob1} \dot{s^*}=L(\vec{s},\vec{u},t),
\end{equation}
\begin{equation}
\label{newprob2}\dot{\vec{s}}=r(\vec{s},\vec{u},t),
\end{equation}
\begin{equation}
\label{newprob3} J=s^*(t_f).
\end{equation}
As a result, the optimization index $J$ is determined solely by the state variables at time $t_f$.

\subsection{The procedure of IDP}
The computational procedure of IDP is described in this subsection.

Before calculation, the time interval, $[t_i,t_f]$, should be discretized into $P$ stages with identical length $l$, {\it i.e.}, $Pl=t_f-t_i$, with constant control variables in each stage. Thus, the protocol of control variables can be represented by a $P$-dimensional array, each element of the array stores the values of control variables in the corresponding time stage. In each time stage, an allowable range of control variables should be specified. The allowable range does not change in one cycle of computation (the definition of a computation cycle will be given below).

A reference $P$-dimensional array, $U$, of control variables pre-exists to initiate each cycle. The elements of $U$ are denoted as $U_i$~($i=1,2,\cdots,P$). The allowable range of control variables is determined by $U$, together with the size of the range, $R$. For the case of one-dimensional control, the allowable range in the $i$th time stage could be $[U_i-R,U_i+R]$. Of course, it is not necessary to set the reference control at the center of the allowable range. To start the first computation cycle, it is necessary to specify an initial $P$-dimensional array as the reference and also the initial size $R$ of the allowable range. It should be noted that $R$ could be either independent or dependent on time. We only discuss the former case here, since no substantial difference exists between the two cases.

In calculation, a matrix of state variables should be first constructed as the mesh grid for dynamic programming. For simplicity, we only discuss the case of one-dimensional control in the following.

First, The allowable interval of control variable at each time stage is uniformly divided into $N_s-1$ parts, with the $N_s$ nodes of this partition taken as the allowable values of control. All the largest allowable values in each time stage are grouped as the first $P$-dimensional array of control ({\it i.e.}, the first control protocol). The second largest ones are grouped as the second $P$-dimensional array, etc. Finally, $N_s$ $P$-dimensional arrays of control variable ($N_s$ control protocols) are obtained. Suppose the arrays are labeled with $m$, the elements of one array are labeled with $n$, then all these control values can be put into a $N_s\times{}P$ matrix, with its elements denoted as $u_{mn}$.

\begin{figure}
\includegraphics[width=3.5in]{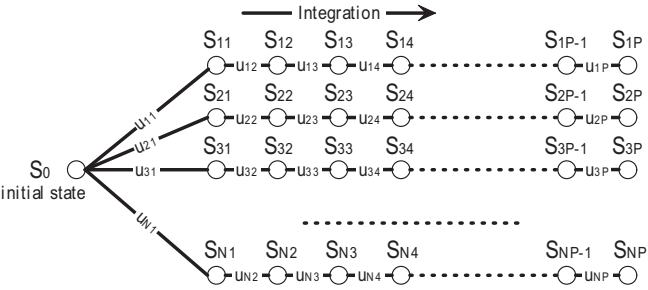}\\
\caption{The process for building the matrix of state variables.}\label{mat}
\end{figure}

Given the matrix of control variable, we can begin to construct the matrix of state variables. The idea is to use each $P$-dimensional array as the control protocol to calculate the corresponding state variables in different time stages. For the arbitrarily selected $j$th array, the calculation starts from the initial state $S_0$, which is determined by the initial conditions. An updated state $S_{j1}$ can be obtained by taking $u_{j1}$ as the control variable and integrating Eq.~(\ref{newprob1}) and Eq.~(\ref{newprob2}) from $t_i$ to $t_i+l$. $S_{j1}$ is stored as an element into the matrix of state variables. After that, starting from $S_{j1}$, another updated state $S_{j2}$ can be obtained by taking $u_{j2}$ as the control variable and integrating from $t_i+l$ to $t_i+2l$. With similar procedure, the $j$th row of the matrix can be constructed by integrating along the $j$th control protocol. After all the integrations are finished, we get a $N_s\times{}P-1$-dimensional matrix of the state variables as shown in Fig.~\ref{mat}. Subsequently, the optimal control value for each element of the matrix can be estimated by the backward iteration method of dynamic programming.

The backward iteration starts from the states $S_{iP-1}$~($i=1,2,\cdots,N_s$), {\it i.e.}, the states at time $t_f-l$. $T$ trial values of control variable should be provided {\it in priori}. Starting from certain $S_{iP-1}$, taking respectively the $T$ trial values as the control variable and integrating Eq.~(\ref{newprob1}) and Eq.~(\ref{newprob2}) from $t_f-l$ to $t_f$, we get $T$ states, $S_{iP-1}^j$~$(j=1,2,\cdots,T)$, at $t=t_f$, which correspond to different $J$ values. The trial control value leading to the minimal $J$ (here we only consider the minimization problems) is chosen as the optimal control variable for $S_{iP-1}$, and will be stored together with $S_{iP-1}$.

Subsequently, for each $S_{iP-2}$~($i=1,2,\cdots,N_s$), by one round of integration from $t_f-2l$ to $t_f-l$, $T$ states, $S_{iP-2}^j$~($j=1,2,\cdots,T$), at $t=t_f-l$ are generated. To proceed, we need to start from these states and integrate from $t_f-l$ to $t_f$. Then, since the resulted $T$ states are at $t=t_f$, their corresponding $J$ values can be compared to judge the optimal control value for $S_{iP-2}$. Similarly with the preceding paragraph, the trial control value leading to the minimal $J$ should be selected as the optimal control of $S_{iP-2}$. The control values of $S_{iP-2}^j$~$(j=1,2,\cdots,T)$ can be determined by the aid of the matrix of state variables. The key is to find the nearest state to $S_{iP-2}^j$ among $S_{iP-1}$~($i=1,2,\cdots,N_s$). If the nearest state to $S_{iP-2}^j$ is $S_{mP-1}$, the control value of $S_{iP-2}^j$ should be set equal to the optimal control of $S_{mP-1}$.

The same procedure applies to other elements $S_{ij}$~($i=1,2,\cdots,N_s.  j=P-3,P-4,\cdots,1$) of the matrix of state variables. One cycle is finished while the above procedure has been performed for $S_0$. After a cycle, we get the matrix with renewed optimal control values $u^*_{ij}$~($i=1,2,\cdots,N_s;j=1,2,\cdots,P-1$) for states $S_{ij}$~($i=1,2,\cdots,N_s;j=1,2,\cdots,P-1$), as well as the optimal control value $u_0^*$ for $S_0$.

Before the next cycle, it is necessary to refresh the reference array of control variable. The procedure is as following. Starting from $S_0$, Eq.~(\ref{newprob1}) and Eq.~(\ref{newprob2}) are first integrated from $t_i$ to $t_i+l$ with the optimal control $u_0^*$, with a new state $S_0^*$ obtained. $u_0^*$ is stored as the first element of a $P$-dimensional array. Then, within the states $S_{1j}$~($j=1,2,\cdots,N_s$), the one nearest to $S_0^*$ is picked out. Suppose the nearest one is $S_{1m}$, starting from $S_0^*$, and integrating Eq.~(\ref{newprob1}) and Eq.~(\ref{newprob2}) from $t_i+l$ to $t_i+2l$ with the optimal control value $u_{1m}^*$ of $S_{1m}$, we get a new state $S_1^*$. Once again, $u_{1m}^*$ is store in the $P$-dimensional array as the second element. A $P$-dimensional array of control variable can be generated by repeating the above procedure step by step. This protocol will be taken as the new reference array of control variable. Besides, the size of the allowable range, $R$, of control variable should be contracted before the new cycle. The new reference array and $R$ together determine the allowable range of control in the next cycle.

A pass is constituted by prescribed number of cycles. Several passes are included in each calculation. At the beginning of a new pass, $R$ is restored partially to its initial value, and the optimal control protocol generated in last pass is used as the initial reference array of control variable.

\subsection{Variable step-length IDP}
The IDP algorithm with variable step-length is almost the same with normal IDP method. Only slight re-formulation is needed. Here the time interval $[t_i,t_f]$ is scaled to $[0,1]$. As in normal IDP method, the $[0,1]$ interval is now uniformly divided into $P$ segments. Furthermore, each segment $i$ corresponds to a real time length of $v_i$, and the real time length in each segment is introduced into IDP algorithm as an extra control variable. Due to the scaling of time, Eq.~(\ref{newprob1}) to Eq.~(\ref{newprob3}) for the $i$th time segment should be expressed as below,
\begin{equation}
\frac{ds^*}{dt'} = L(\vec{s},\vec{u},t')v'_i,
\end{equation}
\begin{equation}
\frac{d\vec{s}}{dt'} = r(\vec{s},\vec{u},t')\frac{dt}{dt'}=r(\vec{s},\vec{u},t')v'_i,
\end{equation}
\begin{equation}
J = s^*(1),
\end{equation}
where, $v'_i$ satisfies
\begin{equation}
\label{renewprob} v'_i \equiv  \frac{dt}{dt'} = v_iP.
\end{equation}
$t'$ is the scaled time. The sum for all $v_i$ should be equal to $t_f-t_i$. This constraint can be incorporated by substituting the origin optimization index $J$ by a new one $I$, which is the sum of $J$ and a penalty function for violating the constraint. For example, $I$ could be chosen as below
\begin{equation}
I=J+\theta(\Sigma_i{}v_i-t_f+t_i)^2.
\end{equation}
Where, $\theta$ is a positive constant, $v_i$ is the value of $v$ at the $i$th time segment. The other kinds of constraints can be implemented in similar way.

\section{Computation Details}
As pointed out in the main text, the time derivative of $\lambda$-protocol is expressed as a temporal function $u$ under current formulation. When searching for the OPMIW, $u$ should be confined within a finite interval at all time. Consequently, larger range of $u$ is necessary to describe the abrupt changing of $\lambda$, {\it e.g.}, a noncontinuous jump or a $\delta$-function trace of $\lambda$ value. However, if in a preliminary calculation the optimized $\lambda$-protocol is found to be monotonic, and some parts of the protocol are jump-like (for example, the $u$ values in these parts are always equal to the allowable upper or lower bound), it is possible to accurately estimate the OPMIW in the following way. Instead of $u=\frac{d\lambda}{dt}$, we can consider the reversed case, $u'=\frac{dt}{d\lambda}$, in calculation because of the monotonicity of $t$ versus $\lambda$, and re-formulate all the functions of $t$ to functions of $\lambda$. Since there is usually no need to specify large range for $u'$, the calculation is simplified. This situation happens to be true for all the overdamped examples studied in current article. Because the initial calculation without restriction almost ensures the monotonicity of $\lambda$, the above-mentioned scheme should be promising for finding the global minimum. For demonstration, the re-formulated Eq.~(\ref{wderiv}), Eq.~(\ref{lamderiv}), Eq.~(\ref{cons1}) and Eq.~(\ref{cons2}) of the overdamped one-spring actuator are shown below,
\begin{equation}
\frac{d{W}_{in}}{d\lambda}=\frac{1}{2}\langle x^2 \rangle,
\end{equation}
\begin{equation}
\frac{dt}{d\lambda}=u',
\end{equation}
\begin{equation}
\frac{d\left\langle{}x\right\rangle}{d\lambda}=fu'-\lambda\left<x\right>u',
\end{equation}
\begin{equation}
\frac{d\left\langle{}x^2\right\rangle}{d\lambda}=2u'-2\lambda\left<x^2\right>u'+2f\left<x\right>u'.
\end{equation}
It is worth noting that although we adopt the formulation with explicitly expressed time derivative of $\lambda$, the other strategies that may be more natural for noncontinuous $\lambda$-protocols, {\it e.g.}, approximating the $\lambda$-protocol with a stepwise function, can also be implemented with IDP.

In current article, we usually study each model (and parameter settings) with two IDP calculations. In the first calculation, the initial reference protocol is selected as a straight line connecting the two points, $(t_i, \lambda_i)$ and $(t_f, \lambda_f)$ (with constant $u\equiv\frac{d\lambda}{dt}$ or $u'\equiv\frac{dt}{d\lambda}$). After that, the resulted control protocol is selected as the initial reference in the second calculation. According to our experience, this procedure can ensure the successful application of IDP in all the examples here.

In computation, we apply the Runge-Kutta method in Ref.~\cite{numrecipe} for integration. The error tolerance in integration is set as $1.e-9$ for the two examples in section \ref{sec:numtest}, and $1.e-7$ for the others. According to our experience, the MIW calculated with $1.e-9$ tolerance are precise up to the seventh digit after the dot, the ones calculated with $1.e-7$ tolerance are precise up to the fifth digit after the dot. Usually, the number of stage, $P$, is selected within the range of $[15, 20]$. In the matrix of state variables, the number of state at each time point is selected within $[10, 20]$. The number of trials along each dimension of control variable is about $10$, leading to around $100$ trials for two-dimensional control variables (including $u\equiv\frac{d\lambda}{dt}$ and the length of time steps). When new cycle begins, the contraction factor for the range of control variables is selected as $0.92$. When new pass begins, the restoration factor for the range is selected as $0.88$. Most of the calculations can be finished within $3$ hours by single-thread computation with Intel(R) Q6700 CPU.

\section*{Acknowledgements} This work is supported by
National Basic Research Program of China (973 Program) under the grant No.~2007CB935903 and NSFC grant No.~10847165.


\newpage

\end{document}